\documentstyle[12pt]{article}
\begin{document}

\def\Jrma{J_{+}^{R}}
\def\Jrme{J_{-}^{R}}
\def\Jrmame{J_{\pm}^{R}}
\def\Jrmema{J_{\mp}^{R}}
\def\Jlma{J_{+}^{L}}
\def\Jlme{J_{-}^{L}}
\def\Jlmame{J_{\pm}^{L}}
\def\Jlmema{J_{\mp}^{L}}
\def\irma{i_{+}^{R}}
\def\irme{i_{-}^{R}}
\def\irmame{i_{\pm}^{R}}
\def\irmema{i_{\mp}^{R}}
\def\ilma{i_{+}^{L}}
\def\ilme{i_{-}^{L}}
\def\ilmame{i_{\pm}^{L}}
\def\ilmema{i_{\mp}^{L}}
\def\jprma{{j^{\psi}}_{+}^{R}}
\def\jprme{{j^{\psi}}_{-}^{R}}
\def\jprmame{{j^{\psi}}_{\pm}^{R}}
\def\jprmema{{j^{\psi}}_{\mp}^{R}}
\def\jplma{{j^{\psi}}_{+}^{L}}
\def\jplme{{j^{\psi}}_{-}^{L}}
\def\jplmame{{j^{\psi}}_{\pm}^{L}}
\def\jplmema{{j^{\psi}}_{\mp}^{L}}
\def\klma{K^{L}_{+}}
\def\klme{K^{L}_{-}}
\def\klmame{K^{L}_{\pm}}
\def\klmema{K^{L}_{\mp}}
\def\krma{K^{R}_{+}}
\def\krme{K^{R}_{-}}
\def\krmame{K^{R}_{\pm}}
\def\krmema{K^{R}_{\mp}}
\def\ylma{Y^{L}_{+}}
\def\ylme{Y^{L}_{-}}
\def\ylmame{Y^{L}_{\pm}}
\def\ylmema{Y^{L}_{\mp}}
\def\yrma{Y^{R}_{+}}
\def\yrme{Y^{R}_{-}}
\def\yrmame{Y^{R}_{\pm}}
\def\yrmema{Y^{R}_{\mp}}
\def\jab{j_{ab}}
\def\pig{\Pi^{g}}
\def\gd{g^{\dagger}}
\def\pipsi{\Pi^{\psi}}
\def\psid{\psi^{\dagger}}
\def\umameg{(1 \pm{\gamma_{5}})}
\def\umemag{(1 \mp{\gamma_{5}})}
\def\l2{\lambda^{2}}
\def\al{\alpha}
\def\fcab{f^{c}_{ab}}
\def\etab{\eta_{ab}}
\def\meio{\frac{1}{2}}
\def\dik{\delta_{ik}}
\def\djl{\delta_{jl}}
\def\dxy{\delta(x_{1}-y_{1})}
\def\umeal{(1-\alpha)}
\def\umaal{(1+\alpha)}
\def\umameal{(1\pm\alpha)}
\def\umemaal{(1\mp\alpha)}
\def\umeal2{(1-\alpha)^{2}}

\title{Current Algebra of Super WZNW Models}

\author{{\normalsize E. Abdalla$\,^1$, M.C.B. Abdalla$\,^2$, O.H.G. Branco$\,^3
$ and L.E. Saltini$\,^4$ \ \ \thanks{Work supported in part by CNPq}}}

 \date{\normalsize
       $^{1,3,4}\,$ Instituto de F\'{\i}sica, Universidade de S\~ao Paulo, \\
       Cx.\ Postal 20516, BR-01498-970 S\~ao Paulo SP / Brazil \\[0.4cm]
       $^2\,$ Instituto de F\'{\i}sica Te\'orica, UNESP, \\
       Rua Pamplona 145, BR-01405-, S\~ao Paulo SP / Brazil}

\maketitle

\begin{abstract}
We derive the current algebra of supersymmetric principal chiral models with a
Wess-Zumino term. At the critical point one obtains two commuting super
Kac-Moody algebra as expected, but in general there are intertwining fields
connecting both right and left sectors, analogously to the bosonic case.
Moreover, in the present supersymmetric extension we have a quadratic algebra,
rather than an affine Lie algebra, due to the mixing between bosonic and
fermionic fields since the purely fermionic sector displays a Lie algebra as
well.
\end{abstract}

 \begin{flushright}
 \parbox{12em}
 { \begin{center}
 Universidade de S\~ao Paulo \\
 IFUSP/P-1026\\
 January 1993
 \end{center} }
 \end{flushright}

\vfill\eject

Since the discovery of higher conservation laws for integrable models,
algebraic methods have been frequently advocated in order to display the
structure of the dynamics of field theory. Higher conserved charges imply non
trivial constraints for correlation functions, and one often finds such a
strong algebraic machinery that a calculable S-matrix turns out to be a
consequence. In the case of conformally invariant field theory, the Virasoro
(and Kac-Moody) constraints fix all correlation functions.

Current algebra for integrable non linear sigma models were, up to very
recently \cite{um} largely unknown. Nevertheless, the Yang-Baxter relations
are the best tools to explore non perturbative properties of integrable models
\cite{dois}. The starting point of the formulation is very generally the
consideration of the Poisson brackets between the spatial part of the Lax pair,
leading to a Lie-Poisson algebra containing an antisymmetric numerical {\bf
r}-matrix  which obeys the Yang-Baxter relation. The Yang-Baxter relation leads
to an (almost) unique S-matrix of the theory \cite{tres,quatro,cinco,dezesseis}
. On the other hand, an algebraic strategy has also been effectively and
successfully used in the case of 2-D conformally invariant theories, in order
to compute correlation functions \cite{sete}, as well as in the case of two
dimensional gravity in the light cone gauge \cite{oito}.

Recently, the non linear $\sigma$-model with a Wess-Zumino term has been
studied in this context \cite{nove}. The current, which corresponds to a piece
of the Lax pair of the model \cite{dez}, is shown to fulfill a new affine
algebra.
Such algebras have played an important role in several cases in field theory
(see refs. \cite{onze} and \cite{doze}). The hope is that after quantization we
could be able to address to the problem of computing exact Greens functions, by
means of higher conserved currents.

In the present paper we consider the supersymmetric Wess-Zumino-Witten
model \cite{treze,catorze}, which is defined by the action

\begin{equation}
\label{um}
S_{susy}(g,\psi)\,=\, \frac{1}{\lambda^{2}} \int d^{2}x\, tr\partial^{\mu}
g^{-1}\partial_{\mu}g\ +
\end{equation}
\[
 + \frac{n}{4\pi}\int^{1}_{0}dr\, \int d^{2}x\,
tr\,g^{-1}_{r}\partial_{r}g\, g^{-1}_{r}
\partial_{\mu}g_{r}\, g^{-1}_{r}\partial_{\nu}g_{r}\ +
\]
\[
+\frac{1}{4\lambda^{2}}\ tr\int d^{2}x \{ \bar{\psi}i\partial\!\!\! / \psi\,
-\,\frac{1}{4}(1-(\frac{n\lambda^{2}}{4\pi})^{2})(\bar{\psi}\psi)^{2}\,+\,
i\frac{n\lambda^{2}}{4\pi}g^{-1}\partial^{\mu}g\, \bar{\psi}\gamma_{5}
\gamma_{\mu}\psi\}
\]

\noindent obtained from the superfield
$G(x,\theta)\,=\,g(x)\,+\,\bar{\theta}\psi(x)\,+\,\frac{1}{2}\bar{\theta}
\theta F(x)$, satisfying the constraint
$G^{\dagger}(x,\theta)G(x,\theta)\,=\,1$, after integrating over the Grassman
variable $\theta$ (see ref \cite{treze} for details, but noticing change in
conventions see below and ref \cite{dezesseis}).

The second term in the right hand side of equation (\ref{um}) is the so called
Wess-Zumino term and only depends linearly on the time derivative of the field
$g$. Hence it proves useful to rewrite it in terms of $A(g)$, introduced in
order to permit the canonical quantization of the theory
 \cite{quinze,dezesseis}, been defined in such a way that

\begin{equation}
\label{dois}
\int^{1}_{0}dr\int d^{2}x \varepsilon^{\mu\nu}\ tr\
g^{-1}_{r}\partial_{r}g_{r}\
g^{-1}_{r}\partial_{\mu}g_{r}\ g^{-1}_{r}\partial_{\nu}g_{r}\ \equiv \ \int
d^{2}x\ tr\ A(g) \partial_{0}g
\end{equation}
and
\begin{equation}
\label{tres}
\frac{\partial A_{ij}}{\partial g_{lk}} \ - \ \frac{\partial A_{kl}}{\partial
g_{ji}} \ = \ \partial_{1} g^{-1}_{il}
g^{-1}_{kj}-g^{-1}_{il}\partial_{1}g^{-1}_{kj} \; .
\end{equation}

We shall use the following notation: $\alpha\,=\,\frac{n \lambda^{2}}{4\pi}$,
$\varepsilon^{01}=1$, $\gamma^{0}=\left(\matrix{0&-i\cr i&0}\right)$,
$\gamma^{1}=\left(\matrix{0&i\cr i&0}\right)$, and $\gamma^{5}\,=
\gamma^{0}\gamma^{1}\,$.
When $\alpha\,=\,\pm 1$ we have a super conformally invariant theory. Canonical
quantization is straightforward, on account of ref \cite{nove} and
\cite{quinze,dezesseis}); we have the following canonically conjugated momenta:

\begin{equation}
\label{quatroa}
\bar{\Pi}^{g}_{ij}=\frac{\partial {\cal L}}{\partial (\partial_{0}
g_{ij})}\,=\,\frac{1}{\lambda^2}\partial^{0}g^{-1}_{ji}\,-\,\frac{n}{4\pi}
A_{ji}(g)\,+\,\frac{i\alpha}{4\lambda^2}\bar{\psi}_{jk}\gamma^1\psi_{km}g^{-1}
_{mi}
\end{equation}
and
\begin{equation}
\label{quatrob}
\Pi^{\psi}_{ij}\,=\, \frac{\partial {\cal L}}
{\partial(\partial_{0}\psi_{ij})}\,=\,\frac{i}{4\lambda^{2}}\bar{\psi}_{ji}
\gamma_{0}\,=\,\frac{i}{4\lambda^{2}}{{\psi}^{\dagger}}_{ji} \; ,
\end{equation}
were the first term in the right side at the equation (\ref{quatroa}),
is the  momentum canonically conjugated to the field $g$ in the principal
chiral model only:

\begin{equation}
\Pi^{g}_{ij}\,=\,\frac{1}{\lambda^{2}} \partial^{0}g^{-1}_{ji}
\label{pige}
\end{equation}
with the following Poisson algebra:

\begin{equation}
\label{cinco}
\{g_{ij}(x),g_{kl}(y)\}\,=\,0 \; ,
\end{equation}
\[ \{\bar{\Pi}^{g}_{ij}(x),\bar{\Pi}^{g}_{kl}(y)\}\,=\,0 \; , \]

\[ \{g_{ij}(x),\bar{\Pi}^{g}_{kl}(y)\}\,=\,\delta_{ik}\delta_{jl}\delta(x_{1}
\,-\, y_{1}) \; , \]

\[ \{\psi_{ij}(x),\psi_{kl}(y)\}_{+}\,=\,0 \; , \]

\[ \{\Pi^{\psi}_{ij}(x),\Pi^{\psi}_{kl}(y)\}_{+}\,=\,0 \; , \]

\[ \{\psi_{ij}(x),{\Pi}^{\psi}_{kl}(y)\}_{+}\,=\,\delta_{ik}\delta_{jl}\delta
(x_{1}-y_{1}) \; . \]

However, as we have already mentioned this model has been obtained via a
superfield formulation where we have imposed the constraint

\begin{equation}
\label{seis}
{G^{\dagger}}(x,\theta)G(x,\theta)\,=\,1
\end{equation}
\noindent which leads, for the fermion component, to the relation
\begin{equation}
\label{sete}
{\psi^{\dagger}}_{ij}\,=\,-g^{-1}_{ik}\psi_{kl}g^{-1}_{lj} \; .
\end{equation}

In the phase space we have the constraint
\begin{equation}
\label{oito}
\Omega_{il}\,=\,\Pi^{\psi}_{li}\,+\,\frac{i}{4\lambda^{2}}g^{-1}_{im}
\psi_{mk}g^{-1}_{kl}
\end{equation}
\noindent which must be implemented using the Dirac method \cite{dezessete}.
The basic element of the method is the dirac matrix
\begin{equation}
\label{novea}
Q\,=\,\{\Omega_{ij}(x),\Omega_{kl}(y)\}\,=\,\frac{i}{2\lambda^{2}}{g^{\dagger}}
_{il}{g^{\dagger}}_{kj} \delta(x_{1}\,-\,y_{1})
\end{equation}
\noindent satisfying
\begin{equation}
\label{noveb}
\int\,dz\{\Omega_{ij}(x),\Omega_{mn}(z)\}\{\Omega_{mn}(z),\Omega_{kl}(y)\}^{-1}
\,=\,\delta_{ik}\delta_{jl}\delta(x_{1}\,-\,y_{1}) \; .
\end{equation}

{}From $Q$ we can ob\-tain the Dirac bra\-ckets (no\-ti\-ce that as we have

\noindent  $\{g_{ij}(x),\Omega_{kl}(y)\}\,=\,0$ the Dirac bracket of $g_{ij}$
with any other field is the Poisson bracket itself). Writing the equation
(\ref{quatroa}) for the field $\Pi^{g}_{ij}$ instead the field
$\bar{\Pi}^{g}_{ij}$, they read:

\begin{equation}
\label{dez}
\{g_{ij}(x),g_{kl}(y)\}_{D}=0 \; ,
\end{equation}

\[ \{g_{ij}(x),\psi_{kl}(y)\}_{D}=0  \; , \]

\[ \{g_{ij}(x),\pipsi_{kl}(y)\}_{D}=0 \; , \]

\[\{g_{ij}(x),\pig_{kl}(y)\}_{D}=\delta_{ik}\delta_{jl}\delta(x_{1}-y_{1})\;
,\]

\[ \{\pig_{ij}(x),\pig_{kl}(y)\}_{D}=
\{ \frac{1}{2}(1-\alpha^{2})[\gd_{jk} \pipsi_{ml}\psi_{mn}\gd_{ni} \ -\
\gd_{li}
\pipsi_{mj}\psi_{mn}\gd_{nk} ] \ + \ \frac{\al}{\l2} F_{ji,lk} \}
\dxy \; , \]

\[ \{\pig_{ij}(x),\psi_{kl}(y)\}_{D}=-\frac{1}{2} \{\dik[\gd_{jm}\psi_{ml}
 -\al \gd_{jm} \gamma_{5}\psi_{ml}] \ +\ \djl [\psi_{km}\gd_{mi} +\al
\psi_{km}\gamma_{5}\gd_{mi}]\}\dxy \; , \]

\[ \{\pig_{ij}(x),\pipsi_{kl}(y)\}_{D}=\meio\{\gd_{li}\pipsi_{kj}+\pipsi_{il}
\gd_{jk} - \al [\pipsi_{il}\gamma_{5}\gd_{jk}-\gd_{li}\gamma_{5}\pipsi_{kj}]
\}\dxy \; , \]

\[ \{\psi_{ij}(x),\psi_{kl}(y)\}_{D}=-\frac{2 \l2}{i} g_{il}g_{kj}\dxy \; , \]

\[ \{ \psi_{ij}(x), \pipsi_{kl}(y)\}_{D}=\meio\dik \djl \dxy \; , \]

\[ \{\pipsi_{ij}(x),\pipsi_{kl}(y)\}_{D}=\frac{-i}{8\l2}\gd_{li}\gd_{jk}\dxy
\; , \]
where
\[ F_{ij,kl}=(\gd_{il})'(x)\gd_{kj}(x)-\gd_{il}(x)(\gd_{kj})'(x) \; .\]

We are now ready to study the relevant algebraic properties of the
supersymetric WZW theory. In analogy with the bosonic case \cite{nove}, we
consider the conserved Noether currents, which in the supersymetric case are
given by:

\begin{equation}
\label{onzea}
{\Jrmame}_{,a}(x)=-\frac{1}{\l2}tr\{(1\mp\al)[g(\dot{(\gd)} \pm (\gd)')+
\frac{i}{4}\psi
\umameg\psid]t_{a}\}(x)
\end{equation}
and
\begin{equation}
\label{onzeb}
{\Jlmame}_{,a}(x)=-\frac{1}{\l2}tr\{(1\pm\al)[(-\dot{(\gd)}\mp(\gd)')g+
\frac{i}{4}\psid
\umameg\psi]t_{a}\}(x) \; .
\end{equation}

Here we introduced the notation with the indices of a basis $t_{a}$ of the Lie
algebra $G$ were the fields $g$ are defined, with the structure constants
$f^{c}_{ab}$ defined as:

\[ [t_{a},t_{b}]\,=\,f^{c}_{ab}\,t_{c} \]

\noindent and any field $J^{R\, or\, L}_{\pm ,a}$ is defined as:

\[ J^{R\, or \, L}_{\pm ,a}\,=\,(J_{\pm},t^{R\, or\, L}_{a})\,=\,-tr(J^{R\, or
\, L}_{\pm}t_{a}) \; .\]

In addition, it is important in the case of theories containing fermions to
introduce the fermionic currents:

\begin{equation}
\label{dozea}
{\irmame}_{,a}(x)=-\frac{i}{4\lambda^{4}} tr[\psi\umemag \psid t_{a}](x)
\end{equation}
and
\begin{equation}
\label{dozeb}
{\ilmame}_{,a}(x)=-\frac{i}{4\lambda^{4}} tr[\psid\umemag\psi t_{a}](x)\; .
\end{equation}

Since we know, from the case of the study of supersymetric non local charges,
that the above purely fermionic objects do also show up independently
 \cite{dezesseis,dezoito,dezenove} of the currents \cite{onze}. Moreover,
there are also intertwining operators already in the bosonic case
\cite{nove,um}, which are described by the fields

\begin{equation}
\label{treze}
j_{ab}(x)=\frac{1}{\l2} tr[\gd t_{a} g t_{b}](x) \; .
\end{equation}

In the case of supersymetric theories we have also the fermionic partner

\begin{equation}
\label{catora}
{\jprmame}_{,ab}(x)=\frac{1}{\l2}tr[\gd\ {\irmame}\ t_{a} g t_{b}](x)
\end{equation}
and
\begin{equation}
\label{catorb}
{\jplmame}_{,ab}(x)=\frac{1}{\l2}tr[\gd t_{a}\ g{\ilmame}\ t_{b}](x)\; .
\end{equation}

We shall see that after coupling bosons with fermions, an infinite number of
fields will be present, as a consequence of a quadratic algebra, which is
absent
in the purely bosonic, as well as in the purely fermionic models. We introduce

\begin{equation}
\label{quinzea}
{\krmame}_{,abc}(x)=-\frac{1}{\l2}tr[\gd t_{a}{\irmame} t_{b} gt_{c}](x)\; ,
\end{equation}

\begin{equation}
\label{quinzeb}
{\klmame}_{,abc}(x)=-\frac{1}{\l2}tr[\gd t_{a}g t_{b} {\ilmame} t_{c}](x) \; ,
\end{equation}

\begin{equation}
\label{quinzec}
{\yrmame}_{,abcd}(x)=\frac{1}{\l2}tr[\gd t_{a} gt_{b} \gd {\irmame} t_{c}
gt_{d}
\; ,
](x)
\end{equation}

\begin{equation}
\label{quinzed}
{\ylmame}_{,abcd}(x)=\frac{1}{\l2}tr[\gd t_{a} gt_{b} \gd t_{c} g {\ilmame}
t_{d}](x) \; .
\end{equation}

We are now in position to write down the full algebra. The purely right sector
is very simple. For the purely left sector one substitutes $(R \rightarrow L)$
and $ \alpha \rightarrow -\alpha $. We have:

\begin{equation}
\label{dezesa}
\{{\Jrmame}_{,a}(x),{\Jrmame}_{,b}(y)\}=-\meio (1\mp\al) \fcab\{(3\pm\al)
{\Jrmame}_{,c}-(1\mp\al){\Jrmema}_{,c} +
\end{equation}
\[ +\frac{\l2}{2}(1\mp\al)[(3\pm 2\al-\al^{2}){\irmame}_{,c} -(5\mp 2\al
+\al^{2}){\irmema}_{,c}]\}\dxy \pm
2(1\mp\al)^{2}\etab\delta'(x_{1}-y_{1})\; , \]

\begin{equation}
\label{dezesb}
\{{\Jrmame}_{,a}(x),{\Jrmema}_{,b}(y)\}=-\meio\fcab\{(1-\al)^{2}{\Jrme}_{,c}
+\umaal^{2}{\Jrma}_{,c} +
\end{equation}
\[ -(1-\al^{2})^{2} \frac{\l2}{2} [{\irme}_{,c}+ {\irma}_{,c}]\}\dxy  \; , \]

\begin{equation}
\label{dezesc}
\{{\irmame}_{,a}(x),{\Jrmame}_{,b}(y)\}=-\meio(1\mp\al)^{2}\fcab{\irmame}_{,c}
\dxy \; ,
\end{equation}

\begin{equation}
\label{dezesd}
\{{\irmame}_{,a}(x),{\Jrmema}_{,b}(y)\}=\meio(1\pm\al)^{2}\fcab{\irmame}_{,c}
\dxy\; ,
\end{equation}

\begin{equation}
\label{dezese}
\{{\irmame}_{,a}(x),{\irmame}_{,b}(y)\}=\frac{1}{\l2}\fcab{\irmame}_{,c}
\dxy \; ,
\end{equation}

\begin{equation}
\label{dezesf}
\{{\irmame}_{,a}(x),{\irmema}_{,b}(y)\}=0 \; ,
\end{equation}
where
\[ \eta_{ab}=(t_{a},t_{b})=-\frac{1}{\l2}tr(t_{a}t_{b}) \]

Notice that the intertwining operators did not appear at all up to this point,
The mixed sector is more involved. We have:

\begin{equation}
\label{dezeta}
\{{\Jrmame}_{,a}(x),{\Jlmame}_{,b}(y)\}=\pm(1-\al^{2})[\umameal j_{ab}(x)
+\umemaal j_{ab}(y)]\delta'(x_{1}-y_{1}) +
\end{equation}
\[ -(1-\al^{2}) \frac{\l2}{4} [(1-\al^{2}) ({\jprme}_{ab}
+{\jprma}_{ab} -{\jplme}_{ab} -{\jplma}_{ab})+8({\jprmema}_{ab}
-{\jplmema}_{ab})]\dxy \]

\begin{equation}
\label{dezetb}
\{{\Jrmame}_{,a}(x),{\Jlmema}_{,b}(y)\}=(1-\al^{2})[\mp(1\mp\al) j'_{ab}(x) +
\end{equation}
\[ -\frac{\l2}{4}\umemaal(3\mp\al)[{\jprme}_{ab}+{\jprma}_{ab}-{\jplme}_{ab}
- {\jplma}_{ab}]]\dxy \]

\begin{equation}
\label{dezetc}
\{{\irmame}_{,a}(x),{\Jlmame}_{,b}(y)\}=\meio(1-\al^{2})[{\jplmame}_{ab}-
{\jprmame}_{ab}]\dxy \; ,
\end{equation}

\begin{equation}
\label{dezetd}
\{{\irmame}_{,a}(x),{\Jlmema}_{,b}(y)\}=\meio\umemaal(3\mp\al)[{\jplmame}_{ab}-
{\jprmame}_{ab}]\dxy \; ,
\end{equation}

\begin{equation}
\label{dezete}
\{{\ilmame}_{,a}(x),{\Jrmame}_{,b}(y)\}=-\meio(1-\al^{2})[{\jplmame}_{ba}-
{\jprmame}_{ba}]\dxy \; ,
\end{equation}

\begin{equation}
\label{dezetf}
\{{\ilmame}_{,a}(x),{\Jrmema}_{,b}(y)\}=-\meio\umameal(3\pm\al)
[{\jplmame}_{ba}- {\jprmame}_{ba}]\dxy \; ,
\end{equation}

\begin{equation}
\label{dezetg}
\{{\irmame}_{,a}(x),{\ilmame}_{,b}(y)\}=\frac{1}{\l2}[{\jplmame}_{ab}-
{\jprmame}_{ab}]\dxy \; ,
\end{equation}

\begin{equation}
\label{dezeth}
\{{\irmame}_{,a}(x),{\ilmema}_{,b}(y)\}=0 \; .
\end{equation}

We see now in equations (\ref{dezeta}, \ref{dezetb}) the explicit
appearance of the bosonic intertwiners as well the fermionic one in equations
(\ref{dezeta}, \ref{dezetb}, \ref{dezetc}, \ref{dezetd}, \ref{dezete},
\ref{dezetf}, \ref{dezetg}). Notice also the
simplicity of the purely fermionic components brackets. In the critical case,
$\alpha \rightarrow +1$ we see that a whole sector decouples completely, the
same being true for $\alpha \rightarrow -1$. We have in that case a super
Kac-Moody algebra, as expected, and the model is completely soluble. In fact,
the conserved charges can be used in order to provide a complete solution of
the Green functions, the only missing ingredient with respect to our
computation being the super Virasoro generators. We finally write down the part
of the algebra involving the intertwiners. We have:

\begin{equation}
\label{dezoa}
\{{\Jlmame}_{,a}(x),j_{bc}(y)\}=-\umameal f^{d}_{ac}j_{bd}\dxy \; ,
\end{equation}

\begin{equation}
\label{dezob}
\{{\Jrmame}_{,a}(x),j_{bc}(y)\}=-\umemaal f^{d}_{ab}j_{dc}\dxy \; ,
\end{equation}

\begin{equation}
\label{dezoc}
\{j_{ab}(x),j_{cd}(y)\}=0 \; ,
\end{equation}

\begin{equation}
\label{dezod}
\{j_{ab}(x),{\ilmame}_{,c}(y)\}=0 \; ,
\end{equation}

\begin{equation}
\label{dezoe}
\{j_{ab}(x),{\irmame}_{,c}(y)\}=0 \; ,
\end{equation}

\begin{equation}
\label{dezof}
\{j_{ab}(x),{\jprmame}_{,cd}(y)\}=0 \; ,
\end{equation}

\begin{equation}
\label{dezog}
\{{\jprmame}_{,ab}(x),{\Jlmame}_{,c}(y)\}=-\meio
\umameal[2{\jprmame}_{\overline{ac},b} -\umameal{\jprmame}_{\overline{ab},c}
- \umemaal{\klmame}_{,abc}]\dxy \; ,
\end{equation}

\begin{equation}
\label{dezoh}
\{{\jprmame}_{,ab}(x),{\Jlmema}_{,c}(y)\}=-\meio
\umemaal[2{\jprmame}_{\overline{ac},b} +\umemaal{\jprmame}_{\overline{ab},c}
- (3\mp\al){\klmame}_{,abc}]\dxy \; ,
\end{equation}

\begin{equation}
\label{dezoi}
\{{\jprmame}_{,ab}(x),{\Jrmame}_{,c}(y)\}=\meio
(1\mp\al)[2{\jprmame}_{\underline{ac},b}
-\umemaal{\jprmame}_{\underline{ca},b}-\umameal{\krmame}_{cab}]\dxy  \; ,
\end{equation}

\begin{equation}
\{{\jprmame}_{,ab}(x),{\Jrmema}_{,c}(y)\}=\meio
(1\pm\al)[2{\jprmame}_{\underline{ac},b}
+\umameal{\jprmame}_{\underline{ca},b}-(3\pm\al){\krmame}_{cab}]\dxy  \; ,
\end{equation}

\begin{equation}
\label{dezoj}
\{{\jprmame}_{,ab}(x),{\ilmame}_{,c}(y)\}=-\frac{1}{\l2}
[{\jprmame}_{\overline{ab},c}-{\klmame}_{abc}]\dxy \; ,
\end{equation}

\begin{equation}
\label{dezok}
\{{\jprmame}_{,ab}(x),{\ilmema}_{,c}(y)\}=0 \; ,
\end{equation}

\begin{equation}
\label{dezol}
\{{\jprmame}_{,ab}(x),{\irmame}_{,c}(y)\}=\frac{1}{\l2}
[{\jprmame}_{\underline{ca},b}-{\krmame}_{cab}]\dxy \; ,
\end{equation}

\begin{equation}
\label{dezom}
\{{\jprmame}_{,ab}(x),{\irmema}_{,c}(y)\}=0\; ,
\end{equation}

\begin{equation}
\label{dezon}
\{{\jprmame}_{,ab}(x),{\jplmame}_{,cd}(y)\}=\frac{1}{\l2}
[{\klmame}_{\underline{cab},d}- {\krmame}_{\overline{cab},d} ]\dxy \; ,
\end{equation}

\begin{equation}
\label{dezoo}
\{{\jprmame}_{,ab}(x),{\jplmema}_{,cd}(y)\}=0\; ,
\end{equation}

\begin{equation}
\label{dezop}
\{{\jprmame}_{,ab}(x),{\jprmame}_{,cd}(y)\}=\frac{1}{\l2}[{\yrmame}_{abcd}
-{\yrmame}_{cdab}]\dxy \; ,
\end{equation}

\begin{equation}
\label{dezoq}
\{{\jprmame}_{,ab}(x),{\jprmema}_{,cd}(y)\}=0\;
\end{equation}
where
\[ ({\jprmame}_{\overline{ab}})_{ij}=\frac{1}{\l2}(\gd {\irmame}
t_{a}gt_{b})_{ij} \]

\[({\jprmame}_{\underline{ab}})_{ij}=\frac{1}{\l2}(\gd {\irmame}
t_{a}t_{b}g)_{ij} \]

\[ ({\krmame}_{\overline{abc}})_{ij}=-\frac{1}{\l2}(\gd t_{a}{\irmame} t_{b} g
t_{c})_{ij} \]

\[ ({\krmame}_{\underline{abc}})_{ij}=-\frac{1}{\l2}(\gd t_{a} t_{b}{\irmame}
t_{c}g)_{ij} \]

\[ ({\jplmame}_{\overline{ab}})_{ij}=\frac{1}{\l2}(\gd t_{a}g{\ilmame}
t_{b})_{ij} \]

\[({\jplmame}_{\underline{ab}})_{ij}=\frac{1}{\l2}(\gd t_{a}t_{b}g{\ilmame}
)_{ij} \]

\[ ({\klmame}_{\overline{abc}})_{ij}=-\frac{1}{\l2}(\gd t_{a}gt_{b}{\ilmame}
t_{c})_{ij} \]

\[ ({\klmame}_{\underline{abc}})_{ij}=-\frac{1}{\l2}(\gd t_{a}t_{b}gt_{c}
{\ilmame})_{ij} \]

It is clear now that the algebra is quadratic. Therefore, a new structure
arises in the case of symmetric theories involving the WZW term. Indeed,
integrability of the bosonic model \cite{dez,vinte} is not clear in the
supersymetric  case \cite{treze}. However it is certainly true that for
$\alpha \,=\,0$ the theory must be integrable. Indeed, there is a considerable
simplification, but the algebra obtained is still quadratic.

\vskip .5cm

Acknowledgement: the authors wish to thank prof. Michael Forger for several
discussions and an initial colaboration, as well as a partial suport of FAPESP.

\end{document}